\documentclass[12pt,a4paper]{iopart}
\usepackage{microtype}
\usepackage{graphics}
\usepackage{iopams}  
\usepackage{amssymb}
\usepackage{amsthm}
\usepackage{cite}

\newcommand{\be}{\begin{equation}}
\newcommand{\ee}{\end{equation}}
\newcommand{\om}{\omega}

\newcommand{\ra}{\rightarrow}
\newcommand{\D}{\mathrm{d}}
\newcommand{\I}{\mathrm{i}}
\newcommand{\E}{\mathrm{e}}
\newcommand{\reals}{\mathbb{R}}
\newcommand{\rh}{\tilde h}
\newcommand{\rs}{\tilde s}
\newcommand{\cE}{\mathcal{E}}
\newcommand{\cM}{\mathcal{M}}
\renewcommand{\Re}{\textrm{Re}}

\newtheoremstyle{myplain}
{5pt}			%Space above 
{5pt}			%Space below 
{\normalsize}	%Body font 
{}			%Indent amount (empty = no indent, \parindent = para indent) 
{\bfseries}		%Thm head font 
{.}			%Punctuation after thm head 
{.5em}		%Space after thm head: " " = normal interword space; \newline = linebreak 
{\thmname{#1}\thmnumber{ #2}\thmnote{~{(#3)}}}

\theoremstyle{myplain}
\newtheorem{theorem}{Theorem}[section]
\newtheorem{example}[theorem]{Example}

\begin{document}

\title{Methods for calculating nonconcave entropies}

\author{Hugo Touchette}

\address{School of Mathematical Sciences, Queen Mary University of London, London E1 4NS, UK}

\begin{abstract}
Five different methods which can be used to analytically calculate entropies that are nonconcave as functions of the energy in the thermodynamic limit are discussed and compared. The five methods are based on the following ideas and techniques: i)~microcanonical contraction, ii) metastable branches of the free energy, iii) generalized canonical ensembles with specific illustrations involving the so-called Gaussian and Betrag ensembles, iv) restricted canonical ensemble, and v) inverse Laplace transform. A simple long-range spin model having a nonconcave entropy is used to illustrate each method.
\end{abstract}


%\maketitle

\section{Introduction}

The recent study of many-body systems interacting via long-range potentials, such as gravitating particles or unscreened plasmas (see \cite{draw2002,campa2008,campa2009,dauxois2010} for other examples), has revealed an interesting property of the entropy that went unnoticed for a surprisingly long time, namely, that it can be \textit{nonconcave} as a function of the energy in the thermodynamic limit. It was known before this discovery that the entropy of finite-size systems could be nonconcave because of boundary or surface contributions (see, e.g., \cite{huller1994b,ispolatov2001b,behringer2006} and \cite{binder1984,challa1986,lee1991} for implicit references to this idea in the context of finite-size scaling). But, in almost all cases, it was assumed that this nonconcavity disappears when taking the thermodynamic limit because the ``bulk'' entropy, which is supposedly always concave as a function of energy, dominates over the ``surface'' entropy. 

We now know that the situation is more complicated and, at the same time, more interesting. If the interaction in a homogeneous many-particle system is \textit{short-range} (see \cite{campa2009} for a definition), then the thermodynamic entropy of this system is essentially always concave as a function of its energy, as has been proved years ago by Ruelle \cite{ruelle1969} (see also Lanford \cite{lanford1973}) using a separation or so-called ``subadditivity'' argument. However, if the interaction is \textit{long-range}, then the subadditivity argument does not work, and the entropy can be either concave or nonconcave \cite{draw2002,campa2008,campa2009}. The latter possibility has for some time been known to arise in mean-field systems, but the crucial and relatively recent input that the study of systems such as gravitating particles has provided is that nonconcave entropies also arise in systems involving physical interactions that are genuinely ``long-range''. In this sense, nonconcave entropies cannot be dismissed as an artifact of mean-field approximations---they are ``physical''. In fact, it is by now established that the nonconcavity of the entropy is related to many interesting physical phenomena, including
\begin{itemize}
\item the existence of energy regions where the heat capacity, defined microcanonically, is negative (something forbidden in the canonical ensemble) \cite{lynden1977,lynden1999,gross2001,draw2002,campa2008,campa2009};
\item the appearance of first-order phase transitions as well as metastable states in the canonical ensemble \cite{bouchet2005,touchette2005a,campa2009};
\item the nonequivalence of the microcanonical and canonical ensembles at the thermodynamic and equilibrium macrostate levels \cite{ellis2000,touchette2004b};
\item a possible ergodicity breaking in microcanonical dynamics \cite{mukamel2005}.
\end{itemize}

We will not be directly concerned here with any of these phenomena; instead, we will consider a problem of a more technical nature having to do with how entropies that are nonconcave can be calculated in practice. Our starting point is the age-old thermodynamic result stating that the entropy of a thermodynamic system is the Legendre transform of its free energy, and vice versa. This duality property can only be true, obviously, if the entropy is concave, since Legendre transforms only yield concave functions. Hence, if one knows or suspects that the entropy of a system is nonconcave, then this entropy cannot be obtained from the canonical ensemble by calculating the Legendre transform of the free energy. How can the entropy be calculated then?

The goal of this paper is to describe and illustrate in the simplest way possible the various methods that have been proposed in the last few years to answer this question. Five methods, which cover the latest studies on the topic of nonconcave entropies, will be covered:
\begin{enumerate}
\item Microcanonical contraction (Sec.~\ref{secmicro});
\item Metastable branches of the canonical free energy (Sec.~\ref{seccrit});
\item Generalized canonical ensembles, with special emphasis on the Gaussian ensemble and a new ensemble called the \textit{Betrag} ensemble (Sec.~\ref{secgen});
\item Restricted canonical ensemble (Sec.~\ref{secres});
\item Inverse Laplace transform (Sec.~\ref{secinv}). 
\end{enumerate}

Each of these methods will be illustrated with a simple spin system introduced in \cite{touchette2003,touchette2008} as a pedagogical model of equilibrium statistical mechanics having a nonconcave entropy. The model, as will become clear, is not meant to represent any real physical system, but has the advantage of being exactly solvable, which makes it useful for demonstrating how the methods listed above work in practice, and for illustrating along the way many general results about nonconcave entropies. 

In theory, all of the methods that will be discussed can be used to calculate any nonconcave entropy, but we will see that some may be more effective or more ``tractable'' than others in practice, depending on the system considered. The question of selecting the ``right'' method for a given system will be discussed at the end of the paper, along with some open problems related to generalized canonical ensembles.

\section{Canonical ensemble}
\label{seccan}

Before we start discussing methods that can be used to calculate nonconcave entropies, let us convince ourselves that the Legendre transform of the canonical free energy does not yield the microcanonical entropy when the latter is nonconcave. This will give us the opportunity to introduce the basic notations used in this paper.

Let $H_N(\om)$ be the Hamiltonian of a classical $N$-particle system, and let $\om$ denote a configuration or \emph{microstate} of this system, and $\Lambda_N$ its configuration space. We define the \emph{thermodynamic free energy} or \emph{free energy density} of the canonical ensemble by the limit
\be
\varphi(\beta)=\lim_{N\ra\infty}-\frac{1}{N}\ln Z_N(\beta),
\label{eqfe1}
\ee
where
\be
Z_N(\beta)=\int_{\Lambda_N} \E^{-\beta H_N(\om)}\, \D \om
\ee
is the $N$-particle partition function. The problem that concerns us here is to determine whether $\varphi(\beta)$ can be used to obtain the \emph{thermodynamic entropy} or \emph{entropy density} of the microcanonical ensemble, defined by
\be
s(u)=\lim_{N\ra\infty}\frac{1}{N}\ln \Omega_N(u),
\label{eqs1}
\ee
where
\be
\Omega_N(u)=\int_{\om\in \Lambda_N: h_N(\om)=u}\D\om=\int_{\Lambda_N}\delta(h_N(\om)-u)\, \D\om 
\label{eqdens1}
\ee
is the density of states, which gives the volume (or, more pictorially, the number) of microstates $\om$ that have a \textit{mean energy} $h_N(\om)=H_N(\om)/N$ equal to $u$. 

As mentioned in the introduction, the common answer to this problem given by most thermodynamics textbooks (see \cite{silhavy1997} for an exception) is that $s(u)$ can \emph{always} be obtained as the Legendre transform of $\varphi(\beta)$, since $\varphi(\beta)$ and $s(u)$ are Legendre transforms of one another. This implies, incidentally, that one can also \emph{always }calculate $\varphi(\beta)$ as the Legendre transform of $s(u)$. But is this really the complete answer? What if $\varphi(\beta)$ is nondifferentiable, as is the case when there is a first-order phase transition in the canonical ensemble? How does one define the Legendre transform for this case? Also, what if $s(u)$ is nonconcave? 

The latter question naturally arises when studying long-range systems. If $s(u)$ is nonconcave, then the Legendre transform of $\varphi(\beta)$ cannot yield $s(u)$ simply because Legendre transforms only yield concave functions \cite{rockafellar1970,tiel1984}. By reading a bit about convex analysis, one learns in fact that the Legendre transform of $\varphi(\beta)$, defined as
\be
\varphi^*(u)=\inf_\beta\{\beta u-\varphi(\beta)\},
\label{eqlf1}
\ee 
yields a concave function corresponding in general to the concave envelope of $s(u)$.\footnote{The transformation defined Eq.~(\ref{eqlf1}) is actually a generalization of the Legendre transform known as the \emph{Legendre-Fenchel} transform, which can be applied to nondifferentiable as well as nonconcave functions; see \cite{rockafellar1970,tiel1984} for more details. In this paper, we refer to this transform as the Legendre transform for simplicity.} Thus, if $s(u)$ is concave, then $s(u)=\varphi^*(u)$, which is to say that $s(u)$ is the Legendre transform of $\varphi(\beta)$. However, if $s(u)$ is nonconcave, then $s(u)\neq\varphi^*(u)$. In this case, only the concave part of $s(u)$ will be recovered from the Legendre transform $\varphi^*(u)$ of $\varphi(\beta)$.

These mathematical results are illustrated in the next example using a simple spin model which will stay with us for the rest of the paper. For proofs of these results, the reader should consult the classical book of Rockafellar \cite{rockafellar1970} or the more readable treatise of van Tiel \cite{tiel1984}.

\begin{figure*}[t]
\centering
\includegraphics{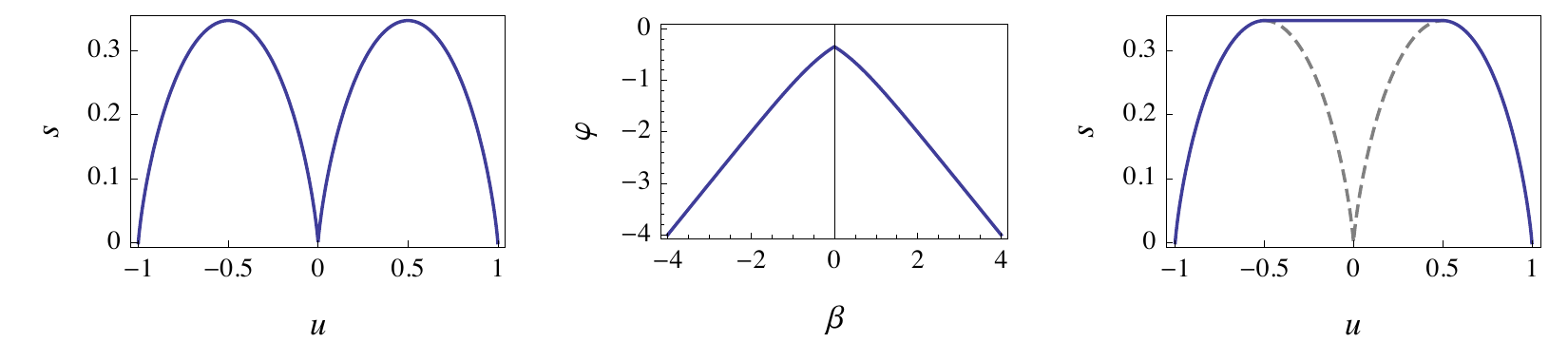}
\caption{Left: Microcanonical entropy $s(u)$ of the block-spin model defined in Example~\ref{ex1}. Center: Canonical free energy $\varphi(\beta)$ of the model. Right: Concave envelope of $s(u)$ (blue) obtained from the Legendre transform of $\varphi(\beta)$.}
\label{figcan1}
\end{figure*}

\begin{example}[Block-spin model~\cite{touchette2008}]
\label{ex1}
Consider the following Hamiltonian:
\be
H_{N}=\frac{N}{2}y +\sum_{i=1}^{N/2} \sigma_i,
\label{eqh1}
\ee
where $y$ and $\sigma_i$, $i=1,2,\dots,N/2$, are spin variables taking values in the set $\{-1,+1\}$. The first term in the Hamiltonian represents the energy of a block of $N/2$ ``frozen'' spins constrained to take the same spin value $y$ ($N$ is assumed to be even). The second term represents the energy of a second block of $N/2$ ``free'' spins which do not interact with each other nor with the first block of spins. 

The entropy density $s(u)$ of this spin system can easily be calculated from the definition of this quantity, i.e., from Eqs.~(\ref{eqs1}) and (\ref{eqdens1}). This calculation is presented in \cite{touchette2008} with the result
\be
s(u)=\frac{1}{2}\left\{ 
\begin{array}{lll}
 s_\sigma(2u+1) &  & u\in [-1,0) \\ 
 s_\sigma(2u-1) &  & u\in [0,1],
\end{array}
\right.
\label{eqs2}
\ee
where
\be
s_\sigma(v)=-\left( \frac{1-v}2\right) \ln \left( \frac{1-v}2\right) -\left( \frac{1+v}2\right) \ln \left( \frac{1+v}2\right)
\ee
is the entropy of the ``free'' spins $\sigma_i$. As is clear from Fig.~\ref{figcan1}, $s(u)$ is a nonconcave function of $u$ as it has more than one maximum\footnote{It is easy to see that a concave function can only have one maximum.} and its derivative is non-monotonic.

To verify that this nonconcave entropy cannot be obtained from $\varphi(\beta)$, we proceed to calculate $\varphi^*(u)$. The direct evaluation of $Z_N(\beta)$ yields for this model
\be
Z_{N}(\beta)=(\E^{\beta N/2}+\E^{-\beta N/2})(\E^{\beta}+\E^{-\beta})^{N/2},
\label{eqpart1}
\ee
so that 
\be
\varphi(\beta)=-\frac{1}{2}|\beta|-\frac{1}{2}\ln (2\cosh\beta).
\label{eqfr1}
\ee
From this expression, plotted in Fig.~\ref{figcan1}, we then compute the Legendre transform defined in Eq.~(\ref{eqlf1}). This calculation can again be found in \cite{touchette2008}; the result is
\be
\varphi^{*}(u)=\frac{1}{2}
\left\{
\begin{array}{lll}
s_\sigma(2u+1)	& & u\in [-1,-\frac{1}{2})\\
\ln2			& & u\in [-\frac{1}{2},\frac{1}{2}]\\
s_\sigma(2u-1)	& & u\in (\frac{1}{2},1].
\end{array}
\right.
\ee
This function is plotted in Fig.~\ref{figcan1}. We see, as announced, that $\varphi^*(u)$ is a concave function corresponding to the concave envelope of $s(u)$. The part of $s(u)$ that coincides with $\varphi^*(u)$ for $u\in [-1,-\frac{1}{2}]\cup[\frac{1}{2},1]$ is called the \emph{concave} parts of $s(u)$, whereas the part such that $s(u)<\varphi^*(u)$, seen for $u\in(-\frac{1}{2},\frac{1}{2})$, is called the \emph{nonconcave} part of $s(u)$.
\end{example}

\begin{figure*}[t]
\centering
\includegraphics{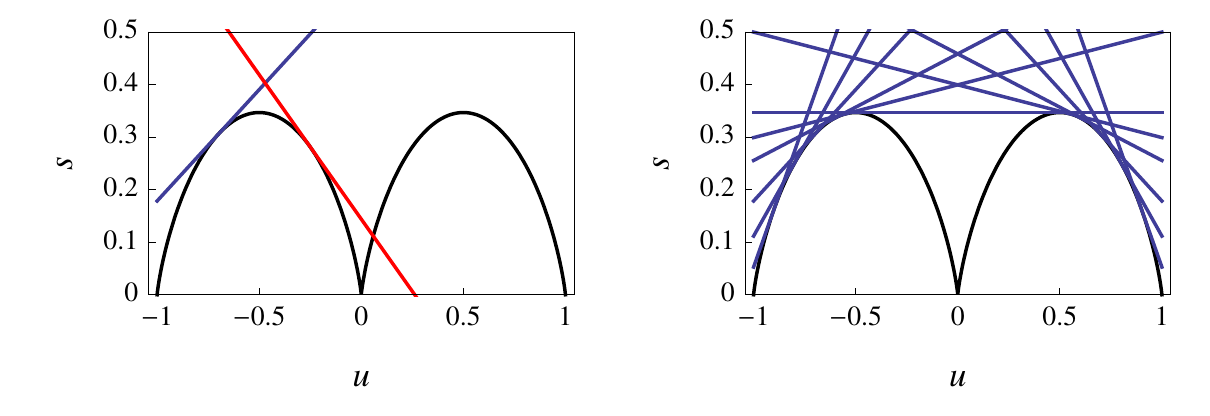}
\caption{(Color online) Left: Illustration of the concept of supporting lines: the line in blue is supporting but not the line in red. Right: The concave envelope of $s(u)$ is given by the set of all supporting lines.}
\label{figsupp1}
\end{figure*}

We can get more insights into the results presented above and illustrated in Fig.~\ref{figcan1} by noting the following extra results of convex analysis:
\begin{itemize}
\item The entropy $s$ at the point $u$ is equal to the Legendre transform of $\varphi$ if and only if one can place a line above the graph of $s(u)$ that touches $s(u)$ without intersect it. When this is possible, we say that $s$ admits a \emph{supporting line} at $u$. Mathematically, this property is expressed as follows: $s=\varphi^*$ at $u$ if, and only if, there exists $\beta\in\reals$ such that
\be
s(v)\leq s(u)+\beta (v-u)
\ee
for all $v$. See \cite{touchette2004b,touchette2009} for more details on the concept of supporting lines.

\item The concave envelope or \emph{concave hull} of $s(u)$ is obtained by constructing the set of all the supporting lines of $s(u)$; see Fig.~\ref{figsupp1}.\footnote{Mathematically, the concave envelope of $s(u)$ is also given by the smallest concave function that majorizes $s(u)$ or by a geometrical construction known as Maxwell's construction; see Sec.~4 of \cite{ellis2004} or Chap.~3 of \cite{touchette2003}.}

\item If $\varphi$ is differentiable at $\beta$, then
\be
s(u_\beta)=\varphi^*(u_\beta)=\beta u_\beta-\varphi(\beta),
\label{eqdlf1}
\ee
where $u_\beta=\varphi'(\beta)$.\footnote{This result is essentially a consequence of the so-called G{\"a}rtner-Ellis Theorem of large deviation theory; see Sec.~5.2 of \cite{touchette2009}.}
\end{itemize}

The first result provides a useful geometric understanding of the points of the entropy that can or cannot be obtained from the Legendre transform of the free energy. This is illustrated in Fig.~\ref{figsupp1}. As for the third result, it shows that $s(u)$ is correctly given by the Legendre transform of $\varphi(\beta)$ for all points $u$ lying in the image of the derivative of $\varphi(\beta)$, i.e., all points $u$ such that $u=\varphi'(\beta)$ for some $\beta\in\reals$. This implies, in particular, that if $\varphi(\beta)$ is everywhere differentiable and the image of $\varphi'$ coincides with the domain of $s$ (i.e., the set of allowed or ``realizable'' values for $H_N/N$), then $s=\varphi^*$ holds globally. We will often come back to this result in the rest of the paper when treating generalized canonical ensembles. 

\section{Microcanonical contraction}
\label{secmicro}

The block-spin model that we have studied in the previous example is simple enough that we can obtain its nonconcave entropy $s(u)$ directly from the definition of this quantity. But, of course, for more realistic and hence more complex models, one should not hope to be able to obtain $s(u)$ in this way. How can $s(u)$ be calculated then?

One answer finds its inspiration from large deviation theory, and attempts to derive $s(u)$ by maximizing another entropy subject to the energy constraint. The basis and hypotheses behind this method are the following \cite{ellis2000}. Given the Hamiltonian $H_N(\om)$, one must be able to find a \emph{macrostate} $M_N(\om)$ such that the following two properties are satisfied:
\begin{itemize}
\item The mean energy $h_N(\om)=H_N(\om)/N$ can re-written as a function of $M_N(\om)$ either exactly or asymptotically in the thermodynamic limit $N\ra\infty$. Mathematically, this implies that there exists a function $\rh(m)$ such that
\be
|h_N(\om)-\rh(M_N(\om))|\ra 0
\ee
uniformily for all $\om\in\Lambda_N$ as $N\ra\infty$. The function $\rh(m)$ is called the \emph{energy representation function}.

\item There exists an entropy function $\rs(m)$ for $M_N(\om)$, which is to say that the following limit exists:
\be
\rs(m)=\lim_{N\ra\infty}\frac{1}{N} \ln \Omega_N(M_N=m),
\ee
where
\be
\Omega_N(m)=\int_{\om\in \Lambda_N: M_N(\om)=m}\D\om=\int_{\Lambda_N}\delta(M_N(\om)-m)\, \D\om
\ee
counts the number of microstates such that $M_N(\om)=m$. The function $\rs(m)$ is called the \emph{macrostate entropy}.
\end{itemize}

When both of these conditions are satisfied, it is relatively easy to show (see \cite{touchette2003,ellis2004}) that
\be
s(u)=\sup_{m:\rh(m)=u} \rs(m).
\label{eqcon1}
\ee
This formula is what we refer to as a \emph{microcanonical contraction}. The word ``contraction'' comes from the fact that this formula can be derived from a result known in large deviation theory as the \emph{contraction principle} \cite{touchette2009}. 

The same result can also be seen as a form of maximum entropy principle expressing $s(u)$ as the constrained maximization of the macrostate entropy $\rs(m)$. It can be shown that the constrained maximizers of $\rs(m)$ such that $\rh(m)=u$ correspond physically to the equilibrium values of $M_N$ in the microcanonical ensemble with mean energy $u$. By denoting the set of such maximizers by $\cE^u$, we can therefore re-express Eq.~(\ref{eqcon1}) as $s(u)=\rs(\cE^u)$.

\begin{example}
\label{exmicro1}
The calculation of the entropy $s(u)$ of the block-spin model via the contraction formula of Eq.~(\ref{eqcon1}) is presented in \cite{touchette2008}. It is easy to see that a natural choice of macrostate for this model is $m=(y,p)$, where $y$ is the spin value of the ``frozen'' block of spins, and $p$ is the proportion of $+1$ spins in the block of ``free'' spins. In terms of this macrostate, we obviously have
\be
\rh(y,p)=\frac{y}{2}+p-\frac{1}{2}.
\ee
The macrostate entropy for this choice of macrostate is, up to a $1/2$ factor, the Boltzmann-Shannon entropy:
\be
\rs(p)=-\frac{1}{2}p\ln p-\frac{1}{2}(1-p)\ln (1-p).
\ee
This entropy does not depend on $y$ because the block of ``frozen'' spins does not contribute to the entropy of the model. We refer the reader to \cite{touchette2008} again for the calculation of $s(u)$ based on this energy representation function and macrostate entropy.
\end{example}

Other examples of calculations of entropies based on the microcanonical contraction formula include the mean-field Blume-Emery-Griffiths model \cite{barre2001,ellis2004}, the mean-field Potts model \cite{costeniuc2005a,barre2005} (see also Example 5.4 of \cite{touchette2009}), the mean-field Hamiltonian model \cite{barre2005,campa2006a}, the so-called mean-field $\phi^4$ model \cite{hahn2005,hahn2006}, as well as a variant of this model having a nonconcave entropy $s(u,m)$ as a function of the energy $u$ and magnetization $m$ \cite{campa2007}. 

From this list and the form of Eq.~(\ref{eqcon1}), one might conclude that this equation is only good for mean-field models, as these are presumably the only models whose Hamiltonian can be re-expressed as a function of some specially chosen macrostates or ``mean-fields''. However, this is not the case. In theory, at least, it is always possible to express the Hamiltonian of any model, including short-range models, as a function of an infinite-dimensional macrostate known as the \emph{empirical process} (see \cite{ellis1985} and Sec.~5.3.4 of \cite{touchette2009}). But given the infinite-dimensional nature of this macrostate, the calculation of $s(u)$ from it is typically impractical if not impossible. For this reason, the microcanonical contraction formula has mostly, if not only, been used in the context of mean-field and long-range models, which are in any case the models for which nonconcave entropies are expected to arise.

\section{Metastable branches of the free energy}
\label{seccrit}

The microcanonical contraction formula discussed in the previous section involves a \emph{constrained} maximization problem which can be transformed, following the theory of Lagrange multipliers, into an \emph{unconstrained} maximization by considering the function
\be
G_\beta(m)= \rs(m)-\beta \rh(m),
\ee
which involves a Lagrange multiplier $\beta$ associated with the constraint $\rh(m)=u$. The question that we ask in this section is: Can we obtain the set of constrained (global) maximizers of $\rs(m)$ with $\rh(m)=u$, which was denoted in the previous section by $\cE^u$, from the set $\cE_\beta$ of (global) unconstrained maximizers of the new function $G_\beta(m)$?

The answer is, no, at least if $s(u)$ is nonconcave. Using techniques similar to those leading to the microcanonical contraction formula, it can indeed be proved that the canonical free energy $\varphi(\beta)$ is given by the set $\cE_\beta$ through the formula
\be
\varphi(\beta)=\inf_{m}\{\beta\rh(m)-\rs(m)\}=-\sup_{m} G_\beta(m)=-G_\beta(\cE_\beta).
\label{eqcon2}
\ee 
Therefore, if we were able to obtain $\cE^u$ from $\cE_\beta$, we would be in a position to obtain $s(u)$ from $\varphi(\beta)$. But we know that this is not possible when $s(u)$ is nonconcave. Hence, $\cE_\beta$ cannot be the same as $\cE^u$ in general.

This result may be surprising but does not contradict the theory of Lagrange multipliers. What this theory actually says is that the global maximizers of $\rs(m)$ subject to the constraint $\rh(m)=u$ are contained in the set of \emph{critical points} of $G_\beta(m)$, which include the global maximizers of $G_\beta(m)$, but also any local maximizers, minimizers and saddle-points that this function may have. The theory simply does not say what the constrained maximizers of $\rs(m)$ correspond to at the level of $G_\beta(m)$. To obtain this information, one must go deeper into the structure of the microcanonical contraction formula, Eq.~(\ref{eqcon1}), and its canonical counterpart, Eq.~(\ref{eqcon2}), to find the following (see \cite{touchette2005a}):
\begin{enumerate}
\item If $s$ is nonconcave at $u$, then the elements of $\cE^u$ correspond either to local minima of $G_\beta(m)$ or saddle-points of this function depending on the local curvature of $s(u)$.
\item If $s$ is concave at $u$, then the elements of $\cE^u$ are also elements of $\cE_\beta$ for some $\beta\in\reals$, which is consistent with the fact that $s=\varphi^*$ in this case.
\end{enumerate}

We reach two conclusions from these results. The first is that the microcanonical and canonical ensembles are equivalent at the level of thermodynamic properties and equilibrium values of macrostates when $s(u)$ is concave \cite{ellis2000,touchette2003,touchette2004b,touchette2009}. The second is more pragmatic: It is possible to obtain $s(u)$ from the knowledge of the critical points of $G_\beta(m)$, but we must consider all the critical points of this function, not just its global maximizers \cite{bouchet2005}. We must locate, in particular, the local maxima of $G_\beta(m)$, which corresponds physically to \emph{metastable} values of $M_N$ in the canonical ensemble, as well as saddle-points of $G_\beta(m)$, which correspond to \emph{unstable} values of $M_N$ in the same ensemble. This conclusion is put to use in the next example.

\begin{example}
\label{exmbfe1}
For the block-spin model, the function $G_\beta(m)$ has the simple form
\be
G_\beta(y,p)=\rs(p)-\beta\,\rh(y,p),\qquad y\in\{-1,1\},\quad p\in[0,1],
\ee
where $\rs(p)$ and $\rh(y,p)$ are the macrostate entropy and energy representation function, respectively, introduced in Example~\ref{exmicro1}. The calculation of the critical points of this form of $G_\beta(y,p)$ can be found in \cite{touchette2008} as well as in Sec.~5.1 of \cite{touchette2003}. For the purpose of this section, there are two points to note about this solution:
\begin{enumerate}
\item $G_\beta(y,p)$ has no saddle-points, but has local maxima, i.e., metastable states, for all $\beta\in\reals$;
\item If we denote the set of metastable points of $G_\beta(y,p)$ for a given $\beta$ by $\cM_\beta$, then $s(u)=\rs(\cM_\beta)$ for $u=\rh(\cM_\beta)\in [-\frac{1}{2},\frac{1}{2}]$.
\end{enumerate}
The last point demonstrates that the metastable states of $G_\beta(y,p)$ can be used, as claimed, to recover nonconcave points of $s(u)$. In fact, for this particular model, $\cM_\beta$ recovers the whole nonconcave region of $s(u)$. The set $\cE_\beta$ of \emph{stable} or \emph{equilibrium} macrostates recovers only the concave parts of $s(u)$.
\end{example}

The nonconcave points of $s(u)$ can also be related, at the thermodynamic level, to metastable \emph{branches} of the canonical free energy function $\varphi(\beta)$ rather than metastable \emph{states} of the canonical ensemble, as was done above. This is illustrated next. 

\begin{figure*}[t]
\centering
\includegraphics{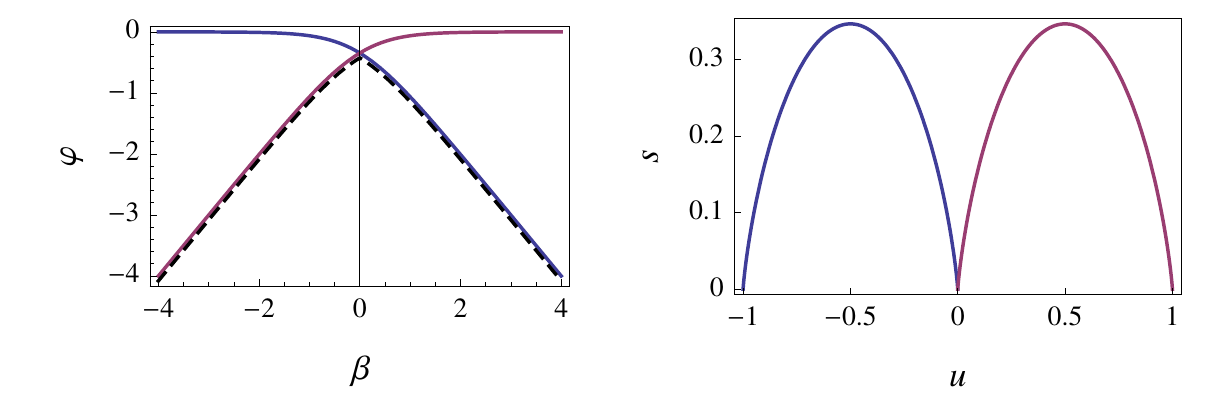}
\caption{(Color online) Left: Plot of $\varphi_1(\beta)$ (blue), $\varphi_2(\beta)$ (purple), and $\varphi(\beta)$ (dashed). The branches of $\varphi_1(\beta)$ and $\varphi_2(\beta)$ that lie above $\varphi(\beta)$ are metastable branches of the free energy. Right: The complete nonconcave entropy is recovered by taking the Legendre transform of $\varphi_1(\beta)$ and $\varphi_2(\beta)$.}
\label{figmeta1}
\end{figure*}

\begin{example}
\label{exmb1}
The exact partition function shown in Eq.~(\ref{eqpart1}) can be put in the form
\be
Z_{N}(\beta)=Z_{N}^{(1)}(\beta)+Z_{N}^{(2)}(\beta),
\label{eqz2}
\ee
where
\be
Z_{N}^{(1)}(\beta)=\E^{\beta N/2}(\E^{\beta}+\E^{-\beta})^{N/2},\qquad Z_{N}^{(2)}(\beta)=\E^{-\beta N/2}(\E^{\beta}+\E^{-\beta})^{N/2}.
\label{eqz3}
\ee
From these two partition functions, it is natural to define two free energy functions, $\varphi_1(\beta)$ and $\varphi_2(\beta)$, using the definition of the free energy shown in Eq.~(\ref{eqfe1}):
\be
\varphi_1(\beta)=-\frac{1}{2}\beta-\frac{1}{2}\ln (2\cosh\beta),\qquad \varphi_2(\beta)=\frac{1}{2}\beta-\frac{1}{2}\ln (2\cosh\beta).
\label{eqfe2}
\ee
The relation between these two free energies and $\varphi(\beta)$ is easily found by using the expression of Eq.~(\ref{eqz2}) in the limit defining $\varphi(\beta)$ to find
\be
\varphi(\beta)=\inf\{\varphi_1(\beta),\varphi_2(\beta)\}=
\left\{
\begin{array}{lll}
\varphi_2(\beta) & & \beta\leq0\\
\varphi_1(\beta) & & \beta>0.
\end{array}
\right.
\ee
This result is illustrated in Fig.~\ref{figmeta1}. The branches of $\varphi_1(\beta)$ and $\varphi_2(\beta)$ that do not contribute to $\varphi(\beta)$ can be interpreted as \emph{metastable} branches of $\varphi(\beta)$, since they continue, in the sense of analytical continuation, the two `stable' branches of $\varphi(\beta)$ while remaining above the `true' minimal equilibrium free energy $\varphi(\beta)$.\footnote{Recall that equilibrium states of the canonical ensemble correspond, according to Gibbs, to those states minimizing the free energy.} As was the case for the metastable states of $G_\beta(y,p)$ studied in the previous example, these metastable branches of $\varphi(\beta)$ completely determine $s(u)$ by Legendre transform:
\be
s(u)=\sup \{\varphi^*_1(u),\varphi_2^*(u)\}=
\left\{
\begin{array}{lll}
\varphi_1^*(u) & & u\in [-1,0)\\
\varphi_2^*(u) & & u\in [0,1].
\end{array}
\right.
\label{eqsm1}
\ee
This result is also illustrated in Fig.~\ref{figmeta1}, and is proved directly by calculating the Legendre transforms of $\varphi_1(\beta)$ and $\varphi_2(\beta)$.
\end{example}

The idea of analytically continuing the free energy around a phase transition point to characterize metastable states was studied for some time in the context of short-range models \cite{langer1967,langer1969,newman1977,newman1980}. However, it was somewhat abandoned after it was realized that continued free energies do not provide correct estimates for the lifetime of metastable states. With hindsight, one could argue that these estimates were wrong because metastables states of short-range systems do not persist in the thermodynamic limit; they arise because of surface effects or, more precisely, because of a ``sub-bulk'' nonconcavity of the entropy, which, as mentioned, disappears in the thermodynamic or ``bulk'' limit. For this reason, metastable states of short-range systems cannot be associated with metastable branches of the free energy because, if they were, then the entropy would have to be nonconcave. 

For long-range systems, the situation is different, since these can have states that are truly metastable in the thermodynamic sense, and are associated with metastable branches of $\varphi(\beta)$, as illustrated by the previous examples. The same phenomenon has also been studied in the context of gravitating systems; see, e.g., \cite{chavanis2006} for a recent review. Still, one must be careful: It is known that different entropies having the same concave envelope lead, by Legendre transform, to the same $\varphi(\beta)$, so it is not possible to uniquely determine $s(u)$ by analytically continuing $\varphi(\beta)$. The existence of metastable branches of $\varphi(\beta)$ must be determined, ultimately, by calculating $Z_N(\beta)$, as was done in the previous example.

\section{Generalized canonical ensembles}
\label{secgen}

The use of generalized ensembles to obtain nonconcave entropies was extensively discussed in previous publications (see \cite{costeniuc2005,touchette2005a,costeniuc2006,touchette2006b,costeniuc2006b}), so we will be brief here. The idea of this method is to obtain $s(u)$ from the Legendre transform of a modified or \emph{generalized free energy function} having the form
\be
\varphi_g(\beta)=\lim_{N\ra\infty}-\frac{1}{N}\ln Z_{N,g}(\beta),
\label{eqgfe1}
\ee
where
\be
Z_{N,g}(\beta)=\int_{\Lambda_N} \E^{-\beta H_N(\om)-Ng(H_N(\om)/N)}\, \D \om
\label{eqgpf1}
\ee
is the generalized partition function. In these expressions, $g$ is a function of the mean energy $H_N/N$, assumed to be continuous. Different choices for this function determine different generalized canonical ensembles that can be used, under some conditions on $g$ (see below), to obtain $s(u)$ even when this function is nonconcave. 

In the following, we will consider two generalized ensembles corresponding to two choices of $g$, and will show that both of these ensembles recover the nonconcave entropy of the block-spin model, but that one is more effective than the other for this purpose. The general result at play behind these two ensembles was proved in \cite{costeniuc2005,costeniuc2006} and can be stated in a simple form as follows: If, for a given choice of function $g$, $\varphi_g(\beta)$ is differentiable at $\beta$, then
\be
s(u)=\beta u-\varphi_g(\beta)+g(u)
\label{eqgenlf1}
\ee
for $u=\varphi_g'(\beta)$. 
This modified Legendre transform, which is written in short as $s=\varphi_g^*+g$, generalizes the standard Legendre transform shown in Eq.~(\ref{eqdlf1}) in an obvious way. In particular, if we are able to find a function $g$ such that $\varphi_g(\beta)$ is everywhere differentiable and the range of $H_N/N$ coincides with the image of $\varphi_g'(\beta)$, then $s=\varphi^*_g+g$ for all $u$ in the domain of $s(u)$. We will see next that such a function can be constructed in some appropriate limit.

\subsection{Gaussian ensemble}

The choice $g(u)=\gamma u^2/2$ with $\gamma\in\reals$ in Eq.~(\ref{eqgpf1}) leads to a generalized ensemble known as the \emph{Gaussian ensemble}. This ensemble
was first introduced in the context of Monte Carlo simulations by Hetherington \cite{hetherington1987a,hetherington1987}, who also discussed its physical interpretation in terms of finite-size heat baths (see \cite{stump1987,challa1988a,challa1988}), and was later re-investigated in the context of nonconcave entropies in \cite{costeniuc2005,costeniuc2006}. It has been applied so far to obtain the nonconcave entropies of two spin models, namely, the mean-field Potts model \cite{costeniuc2006b}, and mean-field Blume-Emery-Griffitths model \cite{frigori2009}. We apply it next to the block-spin model.

\begin{figure*}[t]
\centering
\includegraphics{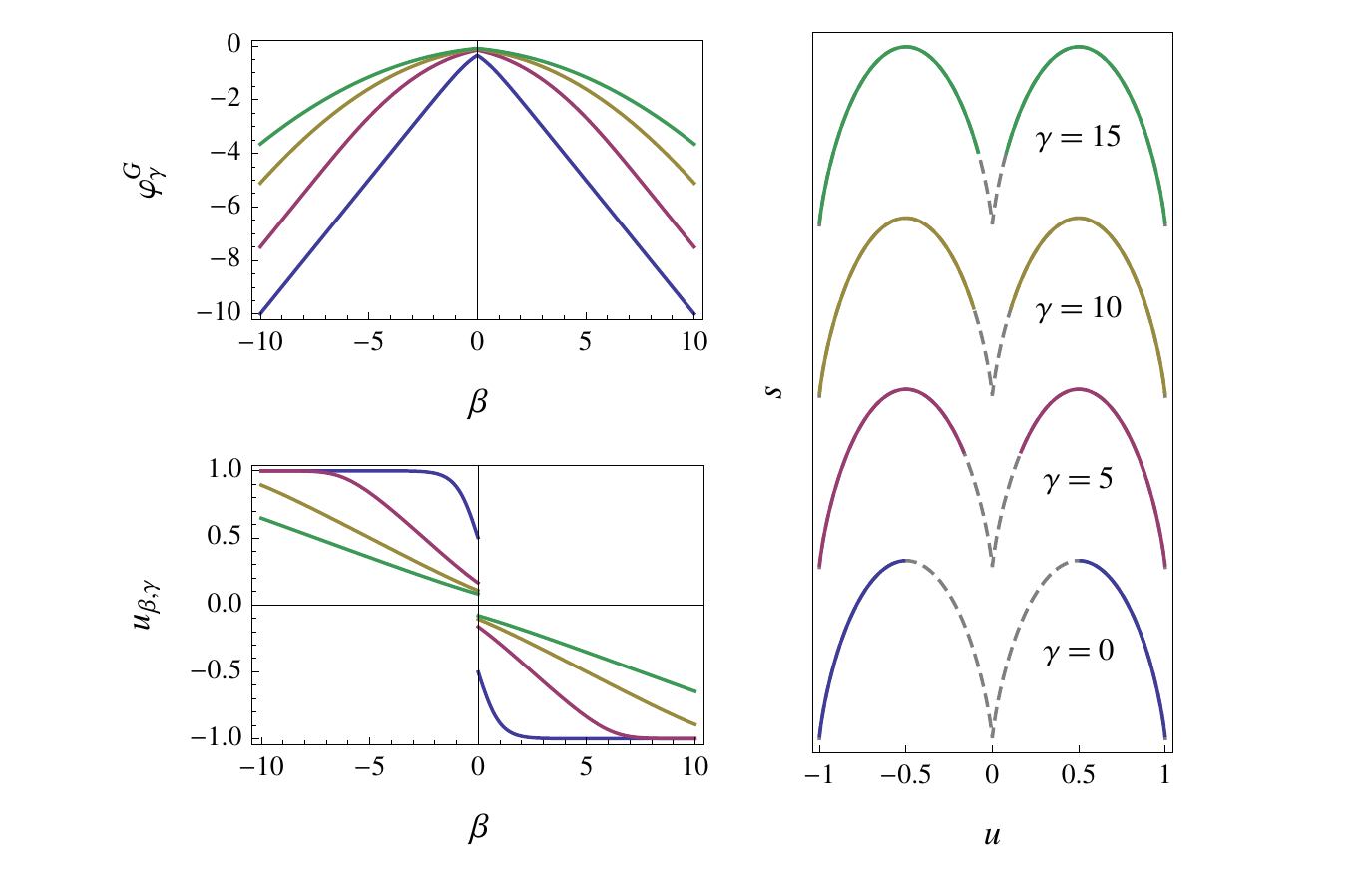}
\caption{(Color online) Gaussian ensemble. Upper left: Gaussian free energy $\varphi^G_\gamma(\beta)$ of the block-spin model for different values of $\gamma$ (see right). Lower left: $\beta$-derivative of $\varphi^G_\gamma(\beta)$. Right: Quadratic Legendre transform of the Gaussian free energy, which recovers the nonconcave entropy $s(u)$ (dashed line) as $\gamma\ra\infty$. The entropy is recovered precisely where $\varphi^G_\gamma$ is differentiable.}
\label{figgauss1}
\end{figure*}

\begin{example}
\label{exgauss1}
The first natural step to take in trying to calculate the generalized partition (\ref{eqgpf1}) with $g(u)=\gamma u^2/2$ is to use the Gaussian integral
\be
\E^{-\gamma u^2/2}=\sqrt{\frac{\gamma}{2\pi}}\int_{-\infty}^\infty \E^{-\gamma t^2/2-\I \gamma u t}\, \D t,
\ee
valid for $\gamma>0$, to obtain
\be
Z_{N,\gamma}^G(\beta)=\sqrt{\frac{\gamma N}{2\pi}}\int_{-\infty}^\infty\D t\, \E^{-\gamma N t^2/2}\, Z_N(\beta+\I\gamma t)
\label{eqgauss1}
\ee
for the Gaussian partition function. Unfortunately, the resulting integral over $t$ cannot be evaluated in general, and in particular not for the block-spin model, despite the simplicity of this model. This point will be discussed in more detail in the concluding section. A similar integral can be obtained for $\gamma<0$, and this one can actually be evaluated using a saddle-point approximation, but, as will be discussed below, the case $\gamma<0$ is not useful for obtaining nonconcave entropies.

It is possible in the end to calculate the Gaussian free energy $\varphi_\gamma(\beta)$ of the block-spin model by generalizing the macrostate representation of $\varphi(\beta)$ found in Eq.~(\ref{eqcon2}) to the Gaussian ensemble:
\be
\varphi_\gamma^G(\beta)=\inf_{y,p}\left\{\beta\rh(y,p)+\frac{\gamma}{2}\rh(y,p)^2-\rs(p)\right\}.
\ee
The minimization problem involved in this expression is easily solved. For $\gamma>0$ and $\beta<0$, the expression between the the curly brackets above is globally minimized for $y=1$ and $p$ solving
\be
\beta+\gamma p-\rs'(p)=0.
\ee
For $\gamma>0$ and $\beta>0$, on the other hand, the same expression is globally minimized for $y=-1$ and $p$ solving
\be
\beta+\gamma(p-1)-\rs'(p)=0.
\ee

Both equations for $p$ are transcendental equations that can be solved numerically to obtain $\varphi_\gamma^G(\beta)$. The result of this numerical calculation is shown in Fig.~\ref{figgauss1}. As can be seen, the Gaussian free energy $\varphi^G_\gamma(\beta)$ obtained for $\gamma>0$ retains the nondifferentiable point of $\varphi(\beta)=\varphi^G_0(\beta)$ at $\beta=0$, but tends to become ``less nondifferentiable'' when $\gamma$ increases, as its left- and right-derivatives approach $0$ for increasing $\gamma$. This behavior of $\varphi^G_\gamma(\beta)$ is illustrated in the lower-left panel of Fig.~\ref{figgauss1}, which shows the plot of $u_{\beta,\gamma}=\partial_\beta\varphi_\gamma^G(\beta)$ for increasing values of $\gamma$. From this plot, we immediately see that the Legendre transform of the Gaussian ensemble, which takes the form
\be
s(u_{\beta,\gamma})=\beta u_{\beta,\gamma}+\frac{\gamma}{2}u_{\beta,\gamma}^2-\varphi_\gamma^G(\beta),
\label{eqglf1}
\ee
should recover more and more nonconcave points of $s(u)$ as $\gamma$ increases. This is confirmed by the right-hand side plot of Fig.~\ref{figgauss1}, which shows the part of the entropy $s(u)$ recovered by Eq.~(\ref{eqglf1}) for $\gamma=0$ (canonical ensemble), $\gamma=5$, $\gamma=10$, and $\gamma=15$. 
\end{example}

\begin{figure*}[t]
\centering
\includegraphics{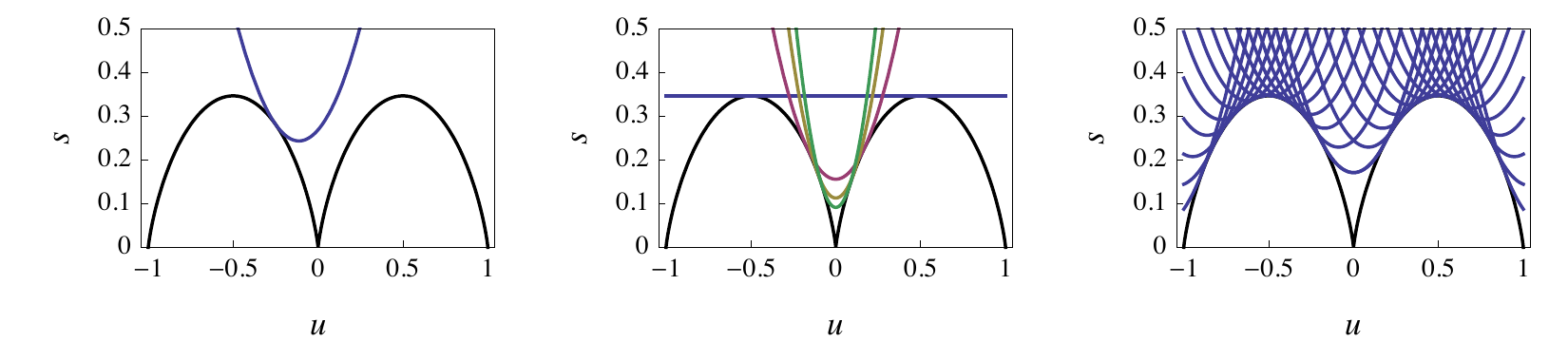}
\caption{(Color online) Left: Illustration of the concept of supporting parabola. Center: Supporting parabolas lying at the center of $s(u)$. The $\gamma$ values of these parabolas are those reported in Fig.~\ref{figgauss1}. The entropy at $u=0$ is recovered only in the limit where $\gamma\ra\infty$ because $s(u)$ has a corner at $u=0$. Right: The parabolic envelope of $s(u)$ is given by the set of all supporting parabolas with given curvature $\gamma$.}
\label{figsupp2}
\end{figure*}

The Gaussian ensemble is an interesting statistical ensemble not only because it can be used to recover nonconcave entropies, as illustrated above, but also because it allows for a natural ``parabolic'' generalization of the concept of supporting lines discussed in Sec.~\ref{seccan}. Because of the quadratic nature of the function $g$ defining this ensemble, it can indeed be proved (see \cite{costeniuc2005,costeniuc2006}) that $s(u)$ is given by the ``quadratic'' Legendre transform of $\varphi_\gamma(\beta)$ shown in Eq.~(\ref{eqglf1}) if
\be
s(v)\leq s(u)+\beta(v-u)+\frac{\gamma}{2}(v-u)^2
\ee 
for all $v$. We say in this case that $s$ admits a \emph{supporting parabola} with curvature $\gamma$ at the point $u$; see Fig.~\ref{figsupp2}. Therefore, the points of $s(u)$ that are recovered by the Gaussian ensemble with parameter $\gamma$ are all (and only) those points that admit a supporting parabola with curvature $\gamma$ or, equivalently, all the points of $s(u)$ coinciding with the \emph{parabolic hull} of this function; see Fig.~\ref{figsupp2}.

We see from this result that the entropy of the block-spin model is obtained only in the limit $\gamma\ra\infty$ because the entropy of this model has a cusp or ``corner'' at $u$, which can only be ``supported'' by a degenerate parabola of infinite curvature, as shown in the center plot of Fig.~\ref{figsupp2}. The same result also explains why the Gaussian ensemble is able to recover nonconcave points of $s(u)$: By modifying the Legendre transform with an added quadratic term, we are able to ``reach'' with a supporting parabola points of $s(u)$ that cannot be ``reached'' with a supporting line; see Fig.~\ref{figsupp2}. This implies, naturally, that the Gaussian ensemble with $\gamma>0$ can only recover more points of $s(u)$ as compared with the canonical ensemble, at least if $s(u)$ has a nonconcave region (and is nondegenerate). (Of course, if $s(u)$ is concave, then the Gaussian ensemble with $\protect{\gamma}>0$ necessarily recovers the whole of $s(u)$ just as the canonical ensemble does.) Conversely, the Gaussian ensemble with $\gamma<0$ must recover \emph{fewer} points of $s(u)$ than the canonical ensemble because points supported by a supporting line may not be supported by a parabola with inverted curvature. This explains our previous observation that the Gaussian ensemble with $\gamma<0$ is not useful for obtaining nonconcave entropies.

\subsection{Betrag ensemble}

We now consider a different ensemble defined by the choice $g(u)=\gamma |u|$ with $\gamma\in\reals$, which will be referred to as the \emph{Betrag ensemble}.\footnote{This ensemble could also be called the ``absolute value ensemble'', but German seems to provide a better name.} This ensemble was mentioned in \cite{costeniuc2006b}, and is somewhat related to piecewise linear Legendre transforms \cite{ellis1995}, but was never applied before to any equilibrium models. 

\begin{figure*}[t]
\centering
\includegraphics{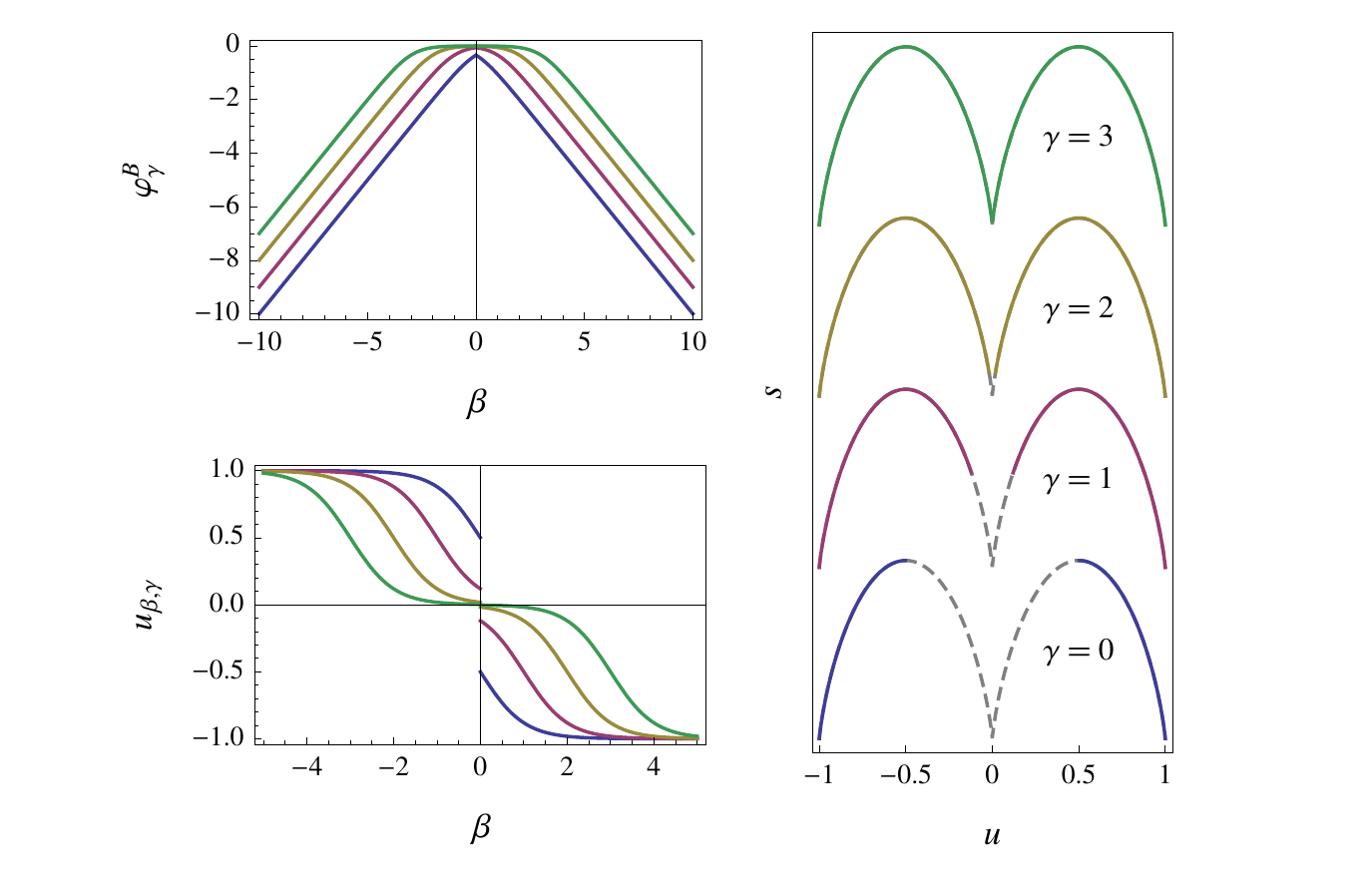}
\caption{(Color online) Betrag ensemble. Upper left: Betrag free energy $\varphi^B_\gamma(\beta)$ of the block-spin model for different values of $\gamma$ (see right). Lower left: Derivative of $\varphi^B_\gamma(\beta)$. Right: Deformed Legendre transform of the Betrag free energy, which recovers the nonconcave entropy $s(u)$ (dashed line) as $\gamma\ra\infty$. The entropy is recovered precisely where $\varphi^B_\gamma$ is differentiable.}
\label{figbet1}
\end{figure*}

\begin{figure}[t]
\centering
\includegraphics{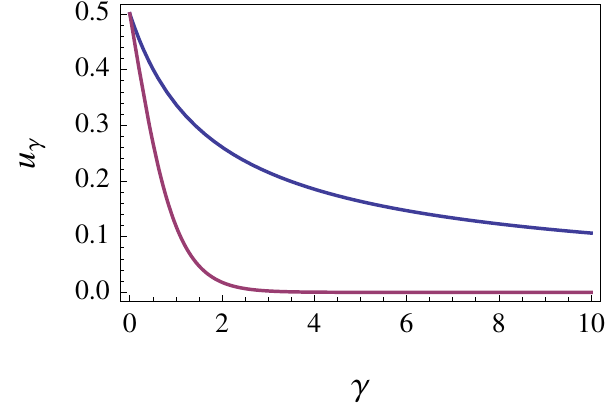}
\caption{(Color online). Comparison of $u_\gamma$ for the Gaussian ensemble (blue line) and Betrag ensemble (purple line).}
\label{figaeq1}
\end{figure}

\begin{example}
\label{exbet1}
The partition function
\be
Z_N^B(\beta)=\int_{\Lambda_N}\E^{-\beta H_N(\om)-\gamma |H_N(\om)|}\, \D\om
\ee
associated with the Betrag ensemble can easily be calculated for the block-spin model because the energy of this model is positive when $y=1$ and negative when $y=-1$. The term $|H_N|$ is easily separable, as a result, and we obtain
\be
Z_{N,\gamma}^B(\beta)=Z_{N}^{(1)}(\beta-\gamma)+Z_{N}^{(2)}(\beta+\gamma),
\ee
where $Z_{N}^{(1)}(\beta)$ and $Z_{N}^{(2)}(\beta)$ are the two canonical partition functions defined in Eq.~(\ref{eqz3}). Given the free energies $\varphi_1(\beta)$ and $\varphi_2(\beta)$ shown in Eq.~(\ref{eqfe2}), we therefore obtain
\be
\varphi^B_\gamma(\beta)=\inf\{\varphi_1(\beta-\gamma),\varphi_2(\beta+\gamma)\}
=\left\{
\begin{array}{lll}
\varphi_1(\beta-\gamma) & & \beta>0\\
\varphi_2(\beta+\gamma) & & \beta\leq 0
\end{array}
\right.
\ee
for the free energy of the Betrag ensemble. 

This free energy function is shown in Fig.~\ref{figbet1}. As for the Gaussian free energy, we see that $\varphi^B_\gamma(\beta)$ has a nondifferentiable point at $\beta=0$ for all the values of $\gamma$ considered, but that the image of the derivative of $\varphi_\gamma^B(\beta)$, which we denote by $u_{\beta,\gamma}$ in the lower-left plot of Fig.~\ref{figbet1}, fills more and more points of the interval $[-1,1]$ as $\gamma$ is increased. These properties of $\varphi^B_\gamma(\beta)$ were also observed for the Gaussian free energy $\varphi^G_\gamma(\beta)$, and imply that the modified Legendre transform of the Betrag ensemble, given by\footnote{See the general result shown in Eq.~(\ref{eqgenlf1}).}
\be
s(u_{\beta,\gamma})=\beta u_{\beta,\gamma}+\gamma |u_{\beta,\gamma}|-\varphi^B_\gamma(\beta),
\ee
recovers more and more points of $s(u)$ as $\gamma$ is increased. This is illustrated in the right-hand side plot of Fig.~\ref{figbet1}. 

As for the Gaussian ensemble, one can show that the Betrag ensemble recovers the full entropy only in the limit $\gamma\ra\infty$. The comparison of Figs.~\ref{figgauss1} and \ref{figbet1} shows, however, that the Betrag ensemble is more efficient at obtaining the full entropy than the Gaussian ensemble. Indeed, both ensembles recover $s(u)$ over an interval of the form $I_\gamma=[-1,-u_\gamma]\cup [u_\gamma,1]$, but $u_\gamma$ converges to $0$ as $\gamma\ra\infty$ faster for the Betrag ensemble than for the Gaussian ensemble; see Fig.~\ref{figaeq1}. This difference in convergence is related to the way the two ensembles achieve equivalence \cite{costeniuc2005}: For the Gaussian ensemble, the limit $\gamma\ra\infty$ is required to recover the whole of $s(u)$ because, as already noted, $s(u)$ has a cusp at $u=0$, whereas for the Betrag ensemble, the limit is needed because $s'(u)$ diverges around its cusp.\footnote{This implies, in particular, that if $s(u)$ has a cusp with finite left- and right-derivatives, then the Betrag ensemble achieves equivalence for a finite value $\gamma>0$, whereas the Gaussian ensemble still requires the limit $\gamma\ra\infty$.}

\end{example}

The possibility of applying the Betrag ensemble to other models rests on being able to separate the partition function of this ensemble into two sums: one involving microstates having a positive energy, and one involving the complementary set of microstates having a negative energy. Such a separation is easily achieved for the block-spin model because of the structure of its Hamiltonian, but one cannot be so optimistic, of course, as to assume that this sort of partitioning trick can be achieved for more realistic models; it all depends on the form of the Hamiltonian $H_N$ that one considers.

In addition to this consideration, it should be clear that, if the nonconcave region of $s(u)$ is located in a region of positive energy, then the Betrag ensemble will not be able to recover or ``express'' any nonconcave points of $s(u)$. In this case, one must replace the function $g(u)=\gamma |u|$ by $g(u)=\gamma |u-u_0|$, where $u_0$ is some fixed value of the mean energy located inside the nonconcave region of $s(u)$. In practice, this means that in order to use the Betrag ensemble in any useful way, one must have some prior information about where $s(u)$ is nonconcave in order to choose the right $u_0$.

\section{Restricted canonical ensemble}
\label{secres}

The concept of restricted canonical ensemble or \emph{restricted partition function} was developed by Penrose and Lebowitz \cite{penrose1979a} not for calculating nonconcave entropies, but as a way to study metastable states in the canonical ensemble. However, from the discussion of these  two topics found in Sec.~\ref{seccrit}, one should expect that restricted partition functions may also be useful for obtaining nonconcave entropies.

The idea behind restricted partition functions is, as the name suggests, to restrict the sum over all microstates $\om\in\Lambda_N$ defining $Z_N(\beta)$ to a subset of $\Lambda_N$ which will be denoted by $R_N$. Thus instead of calculating $Z(\beta)$, one attempts to calculate
\be
Z^R_N(\beta)=\int_{R_N}\E^{-\beta H_N(\om)}\, \D\om.
\ee
The choice of $R_N$ is determined by the fact that, when $s(u)$ is concave, the sum over $\Lambda_N$ in the standard partition function $Z(\beta)$ is dominated in the thermodynamic limit ($N\ra\infty$) by a subset of microstates of $\Lambda_N$ corresponding to the equilibrium states of the canonical ensemble having a fixed energy. However, when $s(u)$ is nonconcave, there is a further important -- yet subdominant -- contribution to the sum of $Z(\beta)$ coming from metastable or unstable states of the canonical ensemble. If one chooses $R_N=\Lambda_N$, then only the dominant equilibrium states will contribute in the partition function. But if one selects $R_N$ so as to exclude the dominant states, then $Z^R_N(\beta)$ will be dominated by the metastable or unstable states. In this case one should be able to recover the nonconcave points of $s(u)$, or at least some of them, by taking the Legendre transform of the free energy function $\varphi^R(\beta)$ associated with $Z^R_N(\beta)$. The next example shows how this works in practice. 

\begin{example}
We have seen with the Betrag ensemble that the microstate space $\Lambda_N$ of the block-spin model can be partitioned, with respect to $H_N$, into microstates of positive and negative energy. Let us use this partition to define a restricted partition function $Z_N^+(\beta)$ by summing only over the microstates having a positive energy:
\be
Z_N^+(\beta)=\sum_{y=1,\sigma_1,\ldots,\sigma_N} \E^{-\beta H_{N}}.
\ee
Going back to the example~\ref{exmb1}, it is easy to see that $Z_N^+(\beta)$ is nothing but the ``metastable'' partition function $Z_{N}^{(2)}(\beta)$ defined in Eq.~(\ref{eqz3}). Therefore, 
\be
\varphi_+^*(u)=\varphi_2^*(u)=\frac{1}{2}s_\sigma(2u-1)
\ee
for $u\in [0,1]$. A similar result can be derived for $u\in [-1,0]$ by calculating a restricted partition function $Z_N^-(\beta)$ for the microstates having a negative energy. In this case, $Z_N^-(\beta)=Z_N^{(1)}(\beta)$ and so
\be
\varphi_-^*(u)=\varphi_1^*(u)=\frac{1}{2}s_\sigma(2u+1)
\ee
for $u\in [-1,0]$. Hence, although neither of the restricted partition functions recovers the whole of $s(u)$, their combination does. 
\end{example}

The difficulty of working with restricted partition functions is similar to that of working with the Betrag ensemble: in both cases, one must be able to calculate a partition function over some restricted set of microstates. Whether this can be done in practice depends on the Hamiltonian $H_N$ considered and, more precisely, on the possibility to use symmetries of this Hamiltonian to partition $\Lambda_N$ in easily-definable sets of microstates. The block-spin model has such a symmetry, as we have seen, which allows for a straightforward calculation of $Z^+_N(\beta)$ and $Z^-_N(\beta)$, as well as $Z_N^B(\beta)$, the betrag partition function. In fact, for this model, the functions $Z^+_N(\beta)$ and $Z^-_N(\beta)$ merely re-create $Z_N^B(\beta)$ but in two separate partition functions instead of one.

Of course, if $H_N$ admits an energy representation function and macrostate entropy function, then restricted partition functions can be calculated for many conceivable restrictions of $\Lambda_N$ simply by restricting the integrals over the macrostate $M_N$ that result from the macrostate representation. In this context, the restriction method can be seen as a way to locate the critical points of the function $G_\beta(m)$, considered in Sec.~\ref{seccrit}, by restricting the range of values allowed for $M_N$.

Finally, note that in the extreme case where the sum-over-states of the standard partition function $Z_N(\beta)$ is restricted only to microstates of constant energy $u$, the resulting restricted free energy function is necessarily equal to the entropy, up to some constant. This follows, of course, because the microcanonical ensemble is a special restricted ensemble that considers only microstates with a constant energy.  

\section{Inverse Laplace transform}
\label{secinv}

The last method that we discuss has been introduced recently in \cite{touchette2010}. Its basis is the inverse Laplace transform that expresses the density of state $\Omega_N(u)$ in terms of the partition function $Z_N(\beta)$:
\be
\Omega_N(u)=\frac{1}{2\pi \I}\int_{r-\I\infty}^{r+\I\infty} Z_N(\beta)\,\E^{\beta Nu}\,\D\beta.
\label{eqilt1}
\ee
This integral is a complex integral along the path or contour $\Re(\beta)=r$, often referred to as the \emph{Bromwich contour}. The value of $r$ used to position this contour must be chosen in the region of convergence of $Z_N(\beta)$, but is otherwise arbitrary.

Since the inverse Laplace transform expresses the density of states exactly in terms of the partition function, it can be used, obviously, to obtain $s(u)$ from $Z_N(\beta)$ even if the former is nonconcave. In fact, it is known that, since the entropy is a thermodynamic-limit function, one needs to know in general only the asymptotic form of $Z_N(\beta)$ as $N\ra\infty$ to obtain $s(u)$ via the inverse Laplace transform. One has to be careful, however, to retain the dominant \emph{and} subdominant terms of $Z_N(\beta)$ when performing any approximations of the Bromwich integral. If one retains only the dominant term, then only the concave envelope of $s(u)$ is recovered, in accordance with our discussion of metastable branches of $\varphi(\beta)$ (see Secs.~\ref{seccrit} and \ref{secres}). This point is illustrated next. 

\begin{example}
Given $\varphi(\beta)$, we can approximate the partition function as $Z_N(\beta)\approx\E^{-N\varphi(\beta)}$, plug this approximation into the integral of the inverse Laplace transform of Eq.~(\ref{eqilt1}), and then naively approximate the resulting integral by its saddlepoint (see, e.g., Appendix C.1 of \cite{touchette2009}) to obtain
\be
\Omega_N(u)\approx \E^{N\inf_\beta\{\beta u-\varphi(\beta)\}}=\E^{N\varphi^*(u)}
\label{eqapp1}
\ee 
and so $s(u)=\varphi^*(u)$. This result is correct, as we know, if $s(u)$ is concave, but not if $s(u)$ is nonconcave; see Sec.~\ref{seccan} and especially Eq.~(\ref{eqlf1}).

A better approximation for $Z_N(\beta)$ is suggested by Eqs.~(\ref{eqz3}) and (\ref{eqfe2}):
\be
Z_N(\beta)\approx \E^{-N\varphi_1(\beta)}+\E^{-N\varphi_2(\beta)}.
\ee
Using this expression in Eq.~(\ref{eqilt1}), we obtain 
\be
\Omega_N(u)\approx \E^{N\varphi_1^*(u)}+\E^{N\varphi_2^*(u)}
\label{eqapp2}
\ee
instead of Eq.~(\ref{eqapp1}), so that
\be
s(u)=\sup \{\varphi^*_1(u),\varphi_2^*(u)\}.
\label{eqsm2}
\ee
We know from Eq.~(\ref{eqsm1}) that this last formula recovers the correct entropy. Therefore, in the case of the block-spin model, the approximation shown in Eq.~(\ref{eqapp2}) is sufficient to obtain $s(u)$. 
\end{example}

The previous example can be generalized to any model whose partition function can be put in the form
\be
Z_N(\beta)=\sum_j c_{N,j}(\beta)\, \E^{-N\varphi_j(\beta)},
\ee
where $\varphi_j(\beta)$ are concave and smooth functions of $\beta$ that do not depend on $N$, and $c_{N,j}(\beta)$ are functions of $\beta$ that  are sub-exponential in $N$. In \cite{touchette2010} it is shown that if these assumptions are satisfied and the coefficients $c_{j,N}(\beta)$ have no poles in the $\beta$-complex plane, then $s(u)$ is given by a direct generalization of Eq.~(\ref{eqsm2}) involving the ``metastable'' free energies $\varphi_j(\beta)$. However, if any of these coefficients have poles in $\beta$, then $s(u)$ is given by a more complicated formula involving the $\varphi_j(\beta)$'s as well as the poles of $c_{N,j}(\beta)$. The surprising effect of these poles is that they determine the presence of linear branches in the graph of $s(u)$, which arise in short-range systems having first-order phase transitions. For more details on these results, the reader is referred again to \cite{touchette2010}.

\section{Comments and open problems}
\label{secconc}

The examples given in the previous sections are very simple, but provide nevertheless a useful guide as to how the different methods that we have covered in this paper can be applied in practice to obtain the nonconcave entropy of more realistic models. They provide, in particular, a good illustration of the properties that one should look for when selecting the right method to use. On the one hand, if the Hamiltonian considered has any symmetries that can be used to partition the microstate space $\Lambda_N$ in easily-definable regions having different energies, then it may be possible to obtain $s(u)$ using the Betrag ensemble or the restricted canonical ensemble. On the other hand, if the Hamiltonian admits a macrostate representation, discussed in Sec.~\ref{secmicro}, then the microcanonical contraction formula or its canonical version involving $G_\beta(m)$ will provide a more direct way to obtain $s(u)$, although all the other methods can also be used in this case, since they all admit a macrostate representation. 

If none of these cases apply, then the application of any of the methods discussed here is likely to lead to difficult or even untractable calculations. This should hardly come as a surprise: after all, the calculation of the standard partition function $Z_N(\beta)$ is known to be a difficult problem in general, and so must be the calculation of any generalizations of $Z_N(\beta)$. In fact, if one cannot analytically calculate $Z_N(\beta)$ for a given model, then it is very unlikely that one will be able to analytical calculate any of the generalized partition functions described before. In this case, one may have to resort to approximation methods or numerical methods, such as Monte Carlo methods based on generalized ensembles (see, e.g., \cite{hetherington1987a,hetherington1987,stump1987,challa1988a,challa1988,neuhaus2006}).

To conclude this paper, we present next a short list of open problems related to the generalized canonical ensembles discussed in Sec.~\ref{secgen}. The first problem is relevant for practical calculations in the Gaussian ensemble, whereas the second is concerned with the numerical implementation of generalized ensembles. The last two problems point to some interesting connections with convex analysis.
\begin{itemize}
\item \emph{Gaussian integral for the Gaussian ensemble}: Study the integral of Eq.~(\ref{eqgauss1}), which expresses the Gaussian partition function in terms of a complex transform of the standard partition function, in order to see if this integral can be approximated in any useful way. We have already commented on the fact this integral cannot be solved for the block-spin model, and is unlikely to be computable in general. The reason for this is that the integrand $\E^{-\gamma N t^2/2}\, Z_N(\beta+\I\gamma t)$ is highly oscillatory when $\gamma>0$, which prevents one from performing any form of saddle-point approximation. Other types of approximation may be possible, however.

\item \emph{Generalized canonical ensembles and multicanonical simulation methods}: There is a strong suggestion that the generalized canonical ensembles discussed in Sec.~\ref{secgen} are related to a set of numerical methods known collectively as \emph{multicanonical} methods or \emph{umbrella sampling} methods (see, e.g., \cite{torrie1974,berg1991,berg1992,lee1993,janke1998b,berg2003,lundow2009}). The exact connection, however, has yet to be made explicit.

\item \emph{Physical interpretation of generalized ensembles}: We mentioned in Sec.~\ref{secgen} that the Gaussian ensemble can be interpreted physically as a statistical-mechanical ensemble describing a sample system coupled to a finite-size heat bath (as opposed to the canonical ensemble which describes a sample system coupled to an \emph{infinite}-size heat bath). Is there a similar physical interpretation for the Betrag ensemble? Are there other generalized ensembles for which a physical interpretation or ``realization'' can be found or constructed?

\item \emph{Supporting functions for generalized canonical ensembles}: We have seen that the concept of supporting lines, which provides a geometrical interpretation of the Legendre transform, is generalized in the Gaussian ensemble to the concept of supporting parabolas. It is not known whether other generalized ensembles admit a similar notion of supporting function. One may ask, for example, whether the supporting function of the Betrag ensemble is the ``absolute value'' function. More generally, are there other types of supporting functions for other choices of $g(u)$?

\item \emph{Moreau transforms}: The quadratic Legendre transform of the Gaussian ensemble, defined in Eq.~(\ref{eqglf1}), appears to be related to a functional transform known in convex analysis as the \emph{Moreau transform} \cite{rockafellar1988}. Are there any known properties of the latter transform that could be used to simplify calculations in the Gaussian ensemble? Moreover, are there generalizations of the Legendre or the Moreau transform that could be used to define other types of generalized canonical ensembles?
\end{itemize}

\section*{Acknowledgments}

I would like to thank Oscar Bandtlow for bits of German offered during tea time, Rosemary J. Harris for reading the manuscript, and an anonymous referee for bringing Ref.~\cite{silhavy1997} to my attention. The hospitality of The Rockefeller University, where most of this paper was written, is also gratefully acknowledged. This work is supported by an RCUK Interdisciplinary Academic Fellowship.

\section*{References}
\bibliography{touchettegenenslanl2}
\bibliographystyle{iopart-num}

\end{document}